\documentclass[aps,prd,showpacs,showkeys,amsmath,amssymb,floatfix,superscriptaddress,twocolumn]{revtex4}
\usepackage{graphicx}
\usepackage{epsfig}
\usepackage{bm}
\usepackage{natbib}
\usepackage{caption}
\usepackage{amsfonts}
\usepackage{xcolor}
\usepackage{changes}

\usepackage{todonotes}
\usepackage[colorlinks=true,
            linkcolor=blue,
            urlcolor=blue,
            citecolor=blue]{hyperref}

\newcommand{\be}{\begin{equation}}
\newcommand{\ee}{\end{equation}}
\newcommand{\bea}{\begin{eqnarray}}
\newcommand{\eea}{\end{eqnarray}}

\usepackage{amsmath, amssymb, graphics, setspace}
\newcommand{\mathsym}[1]{{}}
\newcommand{\unicode}[1]{{}}
\usepackage{textcomp}
\usepackage{eurosym}
\usepackage{amsfonts}
\usepackage{array}
\usepackage{amsthm}
\usepackage{color}
\usepackage{bm}
\usepackage{palatino}

\usepackage{hyperref}

\begin{document}

\title{Cosmology with variable $G$ and nonlinear electrodynamics}

\author{Gabriel W. Joseph}
\email{gabrielwjoseph@gmail.com} \affiliation{Physics Department, Eastern Mediterranean
University, Famagusta, North Cyprus, 99628 via Mersin 10, Turkey}
 
\author{Ali \"{O}vg\"{u}n}
\email{ali.ovgun@emu.edu.tr}
\homepage[]{https://www.aovgun.com}

\affiliation{Physics Department, Eastern Mediterranean
University, Famagusta, North Cyprus, 99628 via Mersin 10, Turkey}
\begin{abstract}
  In a bid to resolve lingering problems in cosmology, more focus is being tilted towards cosmological models in which physical constants of nature are not necessarily real constants but vary with cosmic time. In this paper, we study a cosmological model in nonlinear electrodynamics with Newton's
  gravitational constant $G$, which is not a constant but varies in terms of a power law of
  the scale factor of the universe. Hence, the evolution of the scale factor $a(t)$
  is studied in this model, which depends on a fine-tuning term of nonlinear electrodynamics, $\alpha$. Finally, we verify the stability of the model
  using the speed of sound.
\end{abstract}

\keywords{Cosmology; Nonlinear electrodynamics; Inflation; Acceleration of the universe; Causality; Classical stability; Variable $G$ }
\pacs{95.36.+x, 98.80.-k}
\date{\today}
 \maketitle

\section{Introduction}

Remarkable achievements recorded in the field of cosmology in the past decades
can be attributed to observations of cosmic microwave background (CMB)
radiation and type-Ia supernovae, suggesting that cosmic expansion is
accelerating {\cite{supernova1}}. It is well-known that Maxwell's electrodynamics, as
a source of the Friedmann-Robertson-Walker spacetime, which is the central to the
standard cosmological model (SCM), causes an initial singularity: the
breakdown in geometrical structure of space and time at a finite time in the
past. This initial singularity around the very early universe has revealed an
unpredictable large value of the energy density and curvature of spacetime,
thus departing from the domain of the SCM. This setback has
given rise to other secondary cosmological problems, such as the horizon, flatness, and monopole problems. The secondary problems have
been resolved by using scalar fields. Moreover, magnetic universe models, which
portray no initial singularity due to strong electromagnetic field in the
modified nonlinear electrodynamics (NED) in the early universe, have been used
in the literature to resolve this puzzle
{\cite{DeLorenci:2002mi,vol1,Kruglov:2015fbl,Ovgun:2016oit,Ovgun:2017iwg,Otalora:2018bso,Sarkar:2020fhp,Golovnev:2008cf,novello1,Novello:2006ng,Kruglov:2017vca,Kruglov:2014hpa,Kruglov:2016cdm,Kruglov:2016lqd,Kruglov:2020aje,Sharif:2017pdd,Campanelli:2007cg}}.
The introduction of cosmological constant $\Lambda$, or scalar field with a
type of potential function or modifying the theory of gravity with the aid of
some function $f (r)$, can effectively explain the acceleration of the
universe {\cite{born1}}.

All these models are not without their own setbacks \cite{Poplawski:2012ab,Poplawski:2014dea,Oztas:2018jsu,Dil:2019qos}. For instance, the
smallness of the cosmological constant when compared with the vacuum energy is
very difficult to explain. Furthermore, the choice of $f(r)$ function in the
modified gravity is not unique. Moreover, inflationary cosmological models
provide convincing solutions to the horizon problem, flatness problem, and
small value of cosmological constant, no model of inflation has been
experimentally confirmed.

 In scalar-tensor theory, much potential leads to cosmic inflation and
acceleration of the universe. When NED is coupled to a gravitational field, it can give
the necessary negative pressure and enhance cosmic inflation. Therefore, it is
interesting to explore a new model of NED with variable gravitational constant
where stochastic magnetic field drives the inflation of the universe.

Magnetic fields at different levels in the past have effects
on various cosmological and astrophysical phenomena observed in the universe.
Throughout the universe, physical processes over a large range of scales are influenced by magnetic fields
in the universe. The present and associate magnetic fields could therefore
arise from the flow of materials in the high-temperature vicinity of the big bang.
This corresponds to the primordial magnetic fields (PMF).
Adequate knowledge of the PMF offers information about the
early universe between the inflationary and recombination eras. Such magnetic
fields originate from cosmic phase transition with the bluer spectrum and vector
potential produced during the inflation era. Stochastic inconsistencies of EM field
in relativistic electron-positron plasma produce PMF from thermal
instabilities in the pre-recombination plasma. The plasma sustains the
magnetic fluctuations in the early universe before the era of Big Bang
nucleosynthesis during the radiation-dominated era.

Using the new type of NED containing a stochastic magnetic background with
nonvanishing $B^2$ gives cosmological models void of initial singularities.
Other cosmological models lacking a primordial singularity, such as non-minimal
coupling, quantum gravity effect, Lagrangian with quadratic term, inflation by
scalar fields, and NED without modification of general relativity, have been
introduced in the literature to solve the puzzle of cosmology and mystery of
inflation
{\cite{muk,cos1,cos2,born1,camara,eliz,coupling,durrer,kunze,Azri:2017uor,Azri:2018qux,Azri:2014apa,Hernandez-Almada:2020ulm,Garcia-Aspeitia:2019yni,Vazquez:2018qdg,Sert:2019qfl,Adak:2016led,Okcu:2020ybv,Cruz:2020hkh,Cruz:2018xzn,Cruz:2017bcv,Cordero:2013pua}}.

Max Born and Leopold Infeld used the idea of Gustav who in 1909 began constructing a pure electromagnetic theory of charged particles to propose a new
theory in 1934: fully relativistic and gauge-invariant nonlinear
electrodynamics {\cite{Born:1934gh}}. Born-Infeld proposed a nonlinear fields
Lagrangian with an interesting attribute of transforming to Maxwell's theory for
low electromagnetic fields. Since there are no new degrees of freedoms, such as
scalar fields or branes, works of cosmology described by NED should have some
interesting features of cosmic importance. The sources of cosmic inflation can
be traced to nonlinear electromagnetic radiation which is explained by modified
Maxwell's equations. When coupled with the gravitational fields, NED may give
negative pressure and also can lead to cosmic inflation {\cite{born1}}. The
evolution of the universe, when explored with a new NED model such that
electromagnetic field coupled to gravitational field prevents cosmic
singularity at the big bang. The electromagnetic and gravitational fields were
very strong during the evolution of the early universe, thereby leading to
quantum correction and giving birth to NED {\cite{camara,durrer,kunze}}. One
can present a model of the universe filled with a Born-Infeld type of nonlinear
electromagnetic field, which is inhomogeneous and anisotropic without
singularity.

Recently, interest is being geared towards cosmological models in which
physical constants of nature are varying with time
{\cite{Barrow:1998df,Albrecht:1998ir,Cohen:1998zx}}. For instance, in the
varying speed of light (VSL) theories, where the speed of light is spatial,
temporal and wavelength-dependent, most pending problems of standard
cosmological models are being resolved without considering inflation
{\cite{Moffat:2014poa,Leszczynska:2017bnj,Camara:2007zz}}. The
VSL theories, which deal with the shortcomings of inflation, have not address issues
related to the initial big bang singularity. The Newtonian gravitational constant
$G$ occurs in the source term of Einstein's field equation of the general theory
of relativity, which is a fundamental equation for developing every model of
cosmology. In Einstein's field equation, $G$ acts as a coupling constant
between the geometry of spacetime and matter. In quantum mechanics, $G$ is essential
in the definition of the Planck constant {\cite{Kantha:2016ylw}}. While in SCM,
$G$ is an invariant quantity. It has been noted that there is significant
evidence that the gravitational constant $G$ can vary in time
{\cite{Gv1}}. Dirac, in 1937, argued that variations in $G$ of about 5 parts in
$10^{11}$~per annum could explain the relatively small strength of the
gravitational force compared to other fundamental forces of nature. In
order to unify gravitation and elementary particle physics, Einstein's theory
with time-varying $G$ is already in the literature
{\cite{weyl,dirac,Leszczynska:2017bnj,Singha:2007ey,ggtjp,Malekjani:2015yzp,Sheykhi:2012zzb,Lu:2014rza,Sharif:2012dr,Lu:2009iv,Sheykhi:2010jn,Dungan:2009fp,Borah:2014gca}}.
Singh {\cite{ggtjp}} has presented a cosmological model with $G \sim a^{2 /
\alpha}$, where $a$ is the scale factor of the universe and $\alpha$ a
constant. When either $G$ or the cosmological constant $\Lambda$ is varied
with time, the Einstein field equation is still preserved. It has been shown
that, the variable Newtonian gravitation constant can account for dark energy and
most of its effects, and current dynamical dark energy models using
time-dependent cosmological constant terms are being considered
{\cite{Dungan:2009fp}}.

In this paper, we study a model involving non-linear
electrodynamics coupled to general relativity, additionally assuming a
time-dependent gravitational constant, and study its cosmological dynamics, in
order to unify the different epochs during the evolution of the universe.
We show that variable $G$ in the study of nonlinear electromagnetic
radiation is a source of inflation in the early universe.

The structure of the paper is as follows: in section II, we briefly introduce the
cosmology of a universe filled with nonlinear magnetic monopole fields. In
section III, we obtain the evolution of the universe filled with nonlinear
magnetic monopole fields and variable gravitational constant. In section IV, we
check the stability of the model and give our conclusion in section V.

\section{Non-linear magnetic monopole fields and cosmology}

In nonlinear electrodynamics, we define the Lagrangian density by
{\cite{novello1}}
\begin{equation}
  \mathcal{L}_{NED} = - \frac{\mathcal{F}^{\alpha}}{4}, \label{EMT}
\end{equation}
where $\alpha$ is the electromagnetic fine structure constant and
$\mathcal{F}$ denotes an invariant quantity known as the Maxwell invariant.
Since the matter part of the Lagrangian is independent of the metric's derivatives, in
tensorial language the matter energy-momentum definition using {\eqref{EMT}}
is given as {\cite{Kruglov:2017vca}}
\begin{equation}
  T_{\mu \nu} = - K_{\mu \lambda} F_{\nu}^{~ \lambda} + g_{\mu \nu}
  \mathcal{L}_{NED}, \label{EMT1}
\end{equation}
with
\begin{equation}
  K_{\mu \lambda} = \frac{\partial \mathcal{L}_{NED}}{\partial \mathcal{F}}
  F_{\mu \lambda},
\end{equation}
where $g_{\mu \nu}$ is the metric of spacetime and the indices $\mu$ and
$\nu$ run from 0 to 3. Here, it is assumed that on the cosmic background, there
exists a dominant stochastic magnetic field whose wavelengths are less than its
curvature. Hence, the mean electromagnetic fields then become the source of
Einstein equations {\cite{tolman}}. The averaged electromagnetic fields are
given by {\cite{DeLorenci:2002mi}}:
\begin{equation}
  \langle E \rangle = \langle B \rangle = 0, \langle E_i B_j \rangle = 0,
\end{equation}
\[ \langle E_i E_j \rangle = \frac{1}{3} E^2 g_{ij}, \langle B_i B_j \rangle =
   \frac{1}{3} B^2 g_{ij}, \]
where $\langle$ $\rangle$ denotes averaging brackets used for taking mean of
volume and the indices $i$ and $j$ run from 1 to 3. The wavelength of
radiation is considered to be lower than the volume and the volume smaller
than the curvature.

However, the case of real nonlinear magnetic monopole is when $E^2 = 0$.
Therefore, as obtained from equation {\eqref{EMT}}, the energy density $\rho =
- T^0_{~ 0}$ and the pressure $p = T^i_{~ i} / 3$ of the nonlinear monopole
magnetic field is {\cite{Ovgun:2016oit}}
\begin{eqnarray}
  &  & \rho_{NED} = -\mathcal{L}_{NED},  \label{rho}\\
  &  & p_{NED} =\mathcal{L}_{NED} - \frac{2 B^2}{3}  \frac{\partial
  \mathcal{L}_{NED}}{\partial \mathcal{F}},  \label{p}
\end{eqnarray}
where the definition of $\mathcal{L}_{NED}$ is given in Eq. {\eqref{EMT}} with $\mathcal{F}=
B^2 / 2$.

From the above equations, we obtain the energy density equation $\rho$ and
pressure $p$ as thus:
\begin{equation}
  \begin{array}{l}
    \rho = \rho_{NED} = \frac{2^{- \alpha} (B^2)^{\alpha}}{4}\\
    p = p_{NED} = \frac{2^{- \alpha}}{12} (B^2)^{\alpha}  \left( 4
    \hspace{0.17em} \alpha - 3 \right) .
  \end{array}
\end{equation}
\section{Cosmology with variable $G$ and nonlinear electrodynamics}\label{vsl}

In varying $G$ theories, the action is still
\begin{equation}
  S = \int dx^4  \left( \sqrt{- g}  \left( \frac{R}{16 \pi G}
  +\mathcal{L}_{NED} \right) \right) \label{svsl} .
\end{equation}
Taking the variation of the action with respect to the metric and ignoring
surface terms leads to
\begin{equation}
  G_{\mu \nu} - g_{\mu \nu} \Lambda = \frac{8 \pi G}{c^4} T_{\mu \nu}.
\end{equation}
In cosmological context, the Friedmann Robertson Walker metric is written as
\begin{equation}
  ds^2 = - c^2 dt^2 + a (t)^2  \left[ \frac{dr^2}{1 - Kr^2} + r^2 d \Omega
  \right], \label{dsvsl}
\end{equation}
where $a (t)$ is the scale factor, $t$ is the comoving time and $K = 0, 1, - 1$
represent a flat, closed and open FRW universe, respectively.

For the case of flat FRW $(K = 0)$ and $c = 1$, the Einsteins equations are
{\cite{camara,durrer,kunze}},
\begin{equation}
  H^2 = \left( \frac{\dot{a}}{a} \right)^2 = \frac{8 \pi G (t)}{3} \rho
  \label{fried1},
\end{equation}
\begin{equation}
  \frac{\ddot{a}}{a} = - \frac{4 \pi G (t)}{3}  (\rho + 3 p) . \label{fried2}
\end{equation}
where $H$ represents the Hubble parameter and dot is the differentiation with
respect to time.

However, the conservation equation that follows from
(\ref{fried1})-(\ref{fried2}) is for time variation in $G (t)$ is now
{\cite{Barrow:1998df,Albrecht:1998ir}}:
\begin{equation}
  \dot{\rho} + 3 \frac{\dot{a}}{a}  (\rho + p) = - \rho \frac{\dot{G}}{G},
  \label{cons11}
\end{equation}
where $\omega = \frac{p}{\rho}$ denotes the equation of state parameter for
the dark energy.

Without diving into the dynamics of variable, we shall use
Barrow's ansatz in which the gravitational constant $G$ varies in form of
the power-law of the scale factor as {\cite{Barrow:1998df,Albrecht:1998ir}}:
\begin{equation}
  G = G_0 a^m, \label{gg}
\end{equation}
where $G_0$ is a positive constant. Since we know that $G$ increases with
time, $m$ must be positive. Furthermore, since $G$ depends on the scale factor
of the universe, its time derivatives $\dot{G} $ must be greater than zero.

From the conservation equation {\eqref{cons11}}, we obtain:
\begin{equation}
  - \frac{\partial \mathcal{L}_{NED}}{2 \partial \mathcal{F}} \cdot \left(
  \frac{\mathrm{d}}{\mathrm{d} t} ((B (t))^2) + 4 \hspace{0.17em} \frac{B
  (t)^2  \dot{a}}{a} \right) - \frac{\mathcal{L}_{NED}  \dot{G}}{G} = 0.
\end{equation}
The solution to the above equation gives an important relation between $B (t)$
and $a (t)$ as follows:
\begin{equation}
  B (t) = a (t)^{- 1 / 2 \hspace{0.17em} \frac{4 \hspace{0.17em} \alpha +
  m}{\alpha}} B_0 .
\end{equation}
Conveniently, when written in terms of the scale factor, the evolution of
energy density and pressure are given by:
\begin{equation}
  \rho = \frac{2^{- \alpha}}{4} \left( \mathit{B_0}^2 a (t)^{\frac{- 4
  \hspace{0.17em} \alpha - m}{\alpha}} \right)^{\alpha}, \label{rho1}
\end{equation}
\begin{equation}
  p = \frac{2^{- \alpha}}{12} \left( \mathit{B_0}^2 a (t)^{\frac{- 4
  \hspace{0.17em} \alpha - m}{\alpha}} \right)^{\alpha}  \left( 4
  \hspace{0.17em} \alpha - 3 \right) .
\end{equation}
Then we have:
\begin{equation}
  \rho + p = \frac{2^{- \alpha}}{3} \left( \mathit{B_0}^2 a (t)^{\frac{- 4
  \hspace{0.17em} \alpha - m}{\alpha}} \right)^{\alpha} \alpha,
\end{equation}
\begin{equation}
  \rho + 3 p = 2^{- 1 - \alpha} \left( \mathit{B_0}^2 a (t)^{\frac{- 4
  \hspace{0.17em} \alpha - m}{\alpha}} \right)^{\alpha}  \left( 2
  \hspace{0.17em} \alpha - 1 \right),
\end{equation}
and the EoS parameter $\omega$ is
\begin{equation}
  \omega = \frac{4}{3} \alpha - 1. \label{om}
\end{equation}
Eq. {\eqref{om}} shows some interesting cases of the universe. It follows that
at $\alpha = 0$, $\omega = - 1$, is a cosmological constant case, at

$\alpha = 0.5$, $\omega = - 1$/3 indicates case of dark energy and at $\alpha
= 1$, $\omega = 1 / 3$ denotes the case of ultra-relativistic. The matter
content of the universe is related to its acceleration equation:
\begin{equation}
  \frac{\ddot{a}}{a} = - \frac{4 \pi G (t)}{3}  (\rho + 3 p) . \label{fried22}
\end{equation}
On checking the singularity in energy density and pressure at $a (t)
\rightarrow 0$ and $a (t) \rightarrow \infty$, we obtain,
\begin{equation}
  \lim_{a (t) \rightarrow 0} \rho (t) = \lim_{a (t) \rightarrow 0} p (t) = 0,
  \label{15}
\end{equation}
\begin{equation}
  ~ ~ \lim_{a (t) \rightarrow \infty} \rho (t) = \lim_{a (t) \rightarrow
  \infty} p (t) = 0. \label{166}
\end{equation}
Using the Einstein's field equation and energy momentum tensor, the Ricci
Scalar $R$ which gives the curvature of spacetime is calculated as follows:
\begin{equation}
  R = 8 \hspace{0.17em} \pi \hspace{0.17em} \mathit{G_0} \hspace{0.17em} a
  (t)^m  (\rho - 3 p) . \label{16r}
\end{equation}
The Ricci tensor squared $R_{\mu \nu} R^{\mu \nu}$and the Kretschmann scalar
$R_{\mu \nu \alpha \beta} R^{\mu \nu \alpha \beta}$ are also obtained
{\cite{Ovgun:2016oit}} as:
\begin{equation}
  R_{\mu \nu} R^{\mu \nu} = (8 \hspace{0.17em} \pi \hspace{0.17em}
  \mathit{G_0} \hspace{0.17em} a (t)^m)^2  (\rho^2 + 3 p^2), \label{rrrr}
\end{equation}
\begin{equation}
  R_{\mu \nu \alpha \beta} R^{\mu \nu \alpha \beta} = (8 \hspace{0.17em} \pi
  \hspace{0.17em} \mathit{G_0} \hspace{0.17em} a (t)^m)^2  \left( \frac{5}{3}
  \rho^2 + 2 \rho p + 3 p^2 \right),
\end{equation}
\begin{equation}
  \lim_{a (t) \rightarrow 0} R (t) = \lim_{a (t) \rightarrow 0} R_{\mu \nu}
  R^{\mu \nu} = \lim_{a (t) \rightarrow 0} R_{\mu \nu \alpha \beta} R^{\mu \nu
  \alpha \beta} = 0. \label{18}
\end{equation}
The nature of the scale factor gives the behaviour of the curvature scalar.
Taking the limit of the above equations as the universe accelerates at $a (t)
\rightarrow 0$, \quad we obtain no singularities in the curvature scalar,
Ricci tensor and the Kretschmann scalar.

\section{The Evolution of the Scale Factor of the Universe}

The first Friedmann equation with variable $G (t)$ for the flat universe is
given by
\begin{equation}
  H^2 = \left( \frac{\dot{a}}{a} \right)^2 = \frac{8 \pi G (t)}{3} \rho .
  \label{fm1}
\end{equation}
When a particle moves in one dimension in a potential $V (a)$, the equation of
motion is
\begin{equation}
  \dot{a}^2 + V (a) = 0.
\end{equation}
The potential function
\begin{equation}
  V (a) = -\frac{1}{3} \hspace{0.17em} 2^{1 - \alpha} \pi \hspace{0.17em} G_0
  \hspace{0.17em} (a (t))^{m + 2} \left( B_0^2 (a (t))^{\frac{- 4
  \hspace{0.17em} \alpha - m}{\alpha}} \right)^{\alpha},
\end{equation}
is negative and possesses a maximum at $a = a_c = - C_1$.

Using the {\eqref{gg}} and {\eqref{rho}}, it becomes
\begin{equation}
  \hspace{0.17em} \frac{- 2^{1 - \alpha} \pi \hspace{0.17em} G_0 
  \hspace{0.17em} a^{- 4 \hspace{0.17em} \alpha + 2} B_0^{2 \hspace{0.17em}
  \alpha} + 3 \hspace{0.17em} \dot{a}^2}{3 a^2} = 0,
\end{equation}
then we find the scale factor $a (t)$ is equal to
\begin{eqnarray}
 a (t) \approx \frac{1}{2} \hspace{0.17em} \sqrt{B_0} 2^{\frac{3}{4}} 2^{\frac{3}{4\alpha}}  \left( G_0  \hspace{0.17em} \alpha^2 (C_2 -
  t)^2 \right)^{\frac{1}{4\alpha}}
  3^{- \frac{1}{4\alpha}} \pi^{\frac{1}{4\alpha}}.
\end{eqnarray}
Equation (33) shows that at $t {\rightarrow} 0$, the scale factor $a(t)
{\rightarrow}$ \begin{eqnarray}
 \frac{1}{2} \hspace{0.17em} \sqrt{B_0} 2^{\frac{3}{4}} 2^{\frac{3}{4\alpha}}  \left( G_0  \hspace{0.17em} \alpha^2 C_2^2 \right)^{\frac{1}{4\alpha}} 3^{- \frac{1}{4\alpha}}
  \pi^{\frac{1}{4\alpha}}. 
\end{eqnarray}
This implies that the size of the universe $a(t)$ was never zero.  
\begin{figure}[h]
  \begin{center}
    {\resizebox{12cm}{!}{\includegraphics{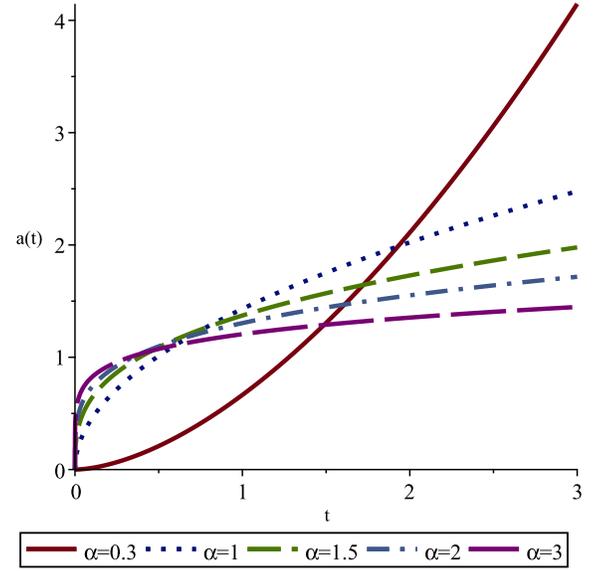}}}
    {\vspace{-15em}}\label{fig1}
  \end{center}
  
  \
  \caption{Plot of the scale factor $a$ versus the time $t$ (for $B_0 = G_0 =
  C_2 = m = 1$).}
\end{figure}

 In the early universe, there is a de Sitter
phase because of the nonlinear corrections to Maxwell's theory, as shown in
Fig.1. Thus,\quad for $\alpha < 1$, the universe enters into its accelerating
expansion phase which indicates the present dark energy dominated epoch. When $\alpha \geq 1$ present a radiation-dominated era with the universe in the phase of decelerating expansion {\cite{Carroll:1997ar}}. 

Introducing the quantity $q$ (the deceleration parameter)
{\cite{Carroll:1997ar}}, we described the expansion of the universe by:
\begin{equation}
  q = - \frac{\ddot{a} a}{(\dot{a})^2} = 9 / 2 \hspace{0.17em} \frac{\rho + 3
  \hspace{0.17em} p}{a \rho} .
\end{equation}
There is a inflation phase for $q < 0$ for $\alpha < 1$ and deceleration phase
for $q > 0$ for $\alpha > 1$ shown in Fig.2.

\begin{figure}[h]
  \begin{center}
    {\resizebox{10cm}{!}{\includegraphics{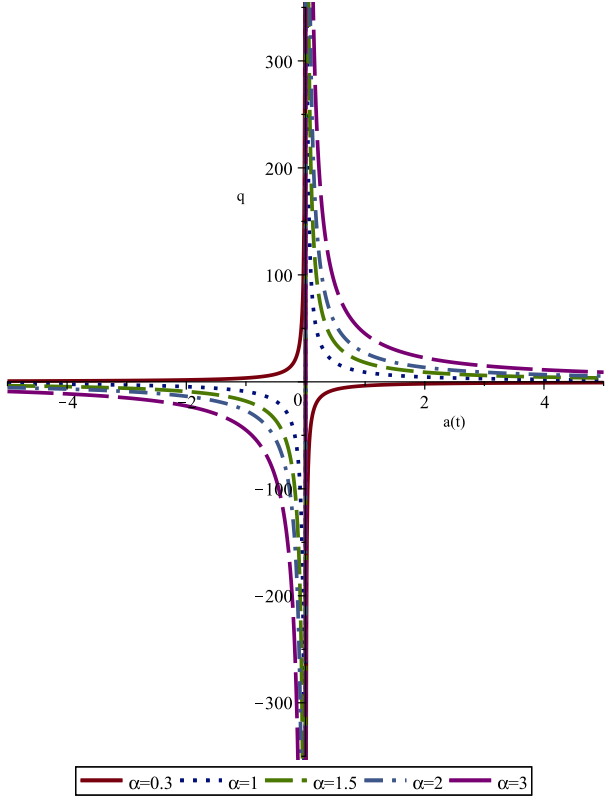}}}\label{fig2} 
  \end{center}
  
  \
  \caption{Plot of the deceleration parameter $q$ versus the scale factor $a$
  (for $B_0 = G_0 = C_2 = m = 1$).}
\end{figure}

To estimate the amount of the inflation, we use the definition of e-foldings
\begin{equation}
  N = \ln \frac{a (t_{end})}{a (t_{in})}
\end{equation}
where $t_{end}$ is the time inflation ends while $t_{in}$ is the time it
begins. For $N \simeq 70$ e-folding, the cosmic flatness and horizon problems
can be resolved. Hence, we obtain the scale factor for beginning time of
inflation (for $m = 1, \alpha = 1, G_0 = 1, B_0 = 1$):
\begin{equation}
  a (t_{in}) = 3.46 \times 10^{- 31} .
\end{equation}
Using the second Friedmann equation Eq. {\eqref{fried22}} which is known as
the acceleration equation for the universe:
\begin{equation}
  \frac{\ddot{a}}{a} = \frac{(1 - 2 \hspace{0.17em} \alpha)}{4 \alpha^2 t^2} .
\end{equation}
It is clear that acceleration stops at $\alpha = 0.5$.

\section{Test of Causality }

For any cosmological model to survive, it is an established that the
speed of sound can not exceed the local speed of light, $c_s \leq 1$. The second requirement for stability is that the square of the speed of sound
must be positive, i.e., $c^2_s > 0$ . In case the model is a classically stable one,
{\cite{sound}} at $E = 0$, we obtain:
\begin{equation}
  c_s^2 = \frac{dp}{d \rho} = \frac{dp / d\mathcal{F}}{d \rho / d\mathcal{F}}
  = - \frac{7}{3} + \frac{4}{3} \alpha .
\end{equation}
A requirement of the classical stability $c^2_s > 0$ is $\alpha > \frac{7}{4}$
and the causality $c_s \leq 1$ is $\alpha \leq \frac{5}{2}$. Hence, the best
value of $\alpha$ for both stability conditions are $\frac{5}{2} \geq \alpha >
\frac{7}{4}$.

\section{Conclusion}

This work studied cosmology with a varying gravitational constant
$G$ and NED in a flat FRW universe. Under changes to the
scale factor, the evolution of the magnetic field reduced to $B (t) = a (t)^{- 1 /
2} B_0$, as obtained in {\cite{Kruglov:2017vca}} where $m = 0$. The evolution of
the scale factor shows that the models give an accelerating, expanding
universe with
\begin{eqnarray}
a (t) \approx \frac{1}{2} \hspace{0.17em} \sqrt{B_0} 2^{\frac{3}{4}} 2^{\frac{3}{4\alpha}}  \left( G_0  \hspace{0.17em} \alpha^2 (C_2 -
  t)^2 \right)^{\frac{1}{4\alpha}} 3^{-\frac{1}{4\alpha}} \pi^{\frac{1}{4\alpha}}, 
\end{eqnarray}
where $B_0$ represents the magnetic induction field at present time $t_0$,
and $\alpha$ is a free parameter presented in Fig. 1. The value of the scale factor obtained indicates that the size of the universe was never zero. By
incorporating NED and variable gravitational constant, we present a model
free from the initial big bang singularity and inflation. As observed in equations {\eqref{15}}, {\eqref{166}}, and {\eqref{18}}, the models portray no
singularity in the energy density, pressure, and curvature terms, respectively.
For $\alpha \geq 1$, the universe is in a radiation-dominated era with the phase of decelerating expansion. However, when
$\alpha = 0.5$, $\omega = - 1$/3 indicates a transition into the present epoch
of dark energy with accelerating expansion. Furthermore, we studied the
stability of this model and observed that it depends of the constant
$\alpha$ and is classically stable for $\frac{5}{2} \geq \alpha >
\frac{7}{4}$. In future studies, we intend to investigate the evolution of the
universe with both varying $G$ and $c$ in NED.

\end{document}